\documentclass
[superscriptaddress,nobibnotes,aps,prd,showpacs,nofootinbib]{revtex4}%
\usepackage{amsmath}
\usepackage{amsfonts}
\usepackage{amssymb}
\usepackage{graphicx}%
\begin{document}
\title{Yukawa-Casimir Wormholes}
\author{Remo Garattini}
\email{remo.garattini@unibg.it}
\affiliation{Universit\`{a} degli Studi di Bergamo, Dipartimento di Ingegneria e Scienze
Applicate,Viale Marconi 5, 24044 Dalmine (Bergamo) Italy and }
\affiliation{I.N.F.N. - sezione di Milano, Milan, Italy.}

\begin{abstract}
In this work, we consider a Yukawa modification of the Casimir wormhole. With
the help of an Equation of State, we impose Zero Tidal Forces. We will examine
two different approaches: in a first approach, we will fix the form of the
shape function of the Casimir wormholes modified by a Yukawa term in three
different ways and finally a superposition of different profiles. In the
second approach, we will consider the original Casimir source modified by a
Yukawa term in three different ways and we will deduce the form of the shape
function In both the approaches the reference energy density will be that of
the Casimir source. Connection with the Absurdly Benign Traversable Wormhole
are also discussed.

\end{abstract}
\maketitle

\section{Introduction}

Yukawa in 1935\cite{Yukawa} proposed to describe nonrelativistic strong
interactions between nucleons with the help of a potential whose profile is%
\begin{equation}
V\left(  r\right)  =-\frac{\alpha}{r}\exp\left(  -\mu r\right)  .\label{(1)}%
\end{equation}
This is nothing but the screened version of the Coulomb potential with
$\alpha$ describing the strength of the interaction and $1/\mu$ its range.
This short range interaction has captured the interest of many researchers who
have adapted it to the Newtonian potential to understand if it has deviations
of the same kind. As a result, the Newtonian gravitational potential between
two point masses $m_{1}$ and $m_{2}$ (atoms for instance) separated by a
distance $r$, acquires a Yukawa correction which formally looks like
Eq.$\left(  \ref{(1)}\right)  $. Indeed, one can write%
\begin{equation}
V\left(  r\right)  =-\frac{Gm_{1}m_{2}}{r}\left(  1+\alpha\exp\left(  -\mu
r\right)  \right)  ,\label{(2)}%
\end{equation}
where $G$ is the gravitational constant. Potentials of the form $\left(
\ref{(2)}\right)  $ have been examined from the astrophysical point of view
with a particular attention also on the graviton mass\cite{StarsGraviton}. It
is interesting to note that Yukawa-type forces are also predicted in the
context of modified gravity theories\cite{MOG} and also in bigravity
theories\cite{Bigravity}. Always in the context of modifying gravity but with
a different framework (MOG), it is possible to obtain black holes and
traversable wormholes\cite{Moffat}. This MOG predicts also a variation of the
Newton's constant $G$, in such a way to obtain a Yukawa term which enters the
metric. Moreover, a Yukawa term seems to be directly involved in the Galaxy
Rotation Curves\cite{MishraSingh}. Even in the context of Casimir effect,
deviations of the Newtonian potential of the form $\left(  \ref{(2)}\right)  $
have been considered\cite{CY}. It is interesting to observe that a connection
between the Casimir forces and the Yukawa profile has been also introduced in
Ref.\cite{Milonni} where Van der Waals himself suggested an interaction
potential of the form%
\begin{equation}
V\left(  r\right)  =-\frac{A}{r}\exp\left(  -Br\right)  ,\label{1}%
\end{equation}
with $A$ and $B$ constants of appropriate dimensions. Since there exists a
connection between the Casimir and the Van der Waals forces in the case of
relatively large separations when the relativistic effects come into play, one
can wonder if Yukawa deformations can play a fundamental r\^{o}le even for
Traversable Wormholes. To further proceed we need to recall the Einstein's
Field Equations (EFE)%
\begin{equation}
G_{\mu\nu}=\kappa T_{\mu\nu}\qquad\kappa=\frac{8\pi G}{c^{4}}%
\end{equation}
in an orthonormal reference frame. In such a frame the EFE reduce to the
following set of equations%
\begin{equation}
\frac{b^{\prime}}{r^{2}}=\kappa\rho\left(  r\right)  ,\label{rho}%
\end{equation}%
\begin{equation}
\frac{2}{r}\left(  1-\frac{b\left(  r\right)  }{r}\right)  \phi^{\prime}%
-\frac{b}{r^{3}}=\kappa p_{r}\left(  r\right)  ,\label{pr0}%
\end{equation}%
\begin{align}
&  \Bigg\{\left(  1-\frac{b}{r}\right)  \left[  \phi^{\prime\prime}%
+\phi^{\prime}\left(  \phi^{\prime}+\frac{1}{r}\right)  \right]  \nonumber\\
&  -\frac{b^{\prime}r-b}{2r^{2}}\left(  \phi^{\prime}+\frac{1}{r}\right)
\Bigg\}=\kappa p_{t}(r),\label{pt0}%
\end{align}
in which $\rho\left(  r\right)  $ is the energy density\footnote{However, if
$\rho\left(  r\right)  $ represents the mass density, then we have to replace
$\rho\left(  r\right)  $ with $\rho\left(  r\right)  c^{2}.$}, $p_{r}\left(
r\right)  $ is the radial pressure, and $p_{t}\left(  r\right)  $ is the
lateral pressure. The EFE $\left(  \ref{rho}\right)  $, $\left(
\ref{pr0}\right)  $ and $\left(  \ref{pt0}\right)  $ have been obtained with
the help of the line element%
\begin{equation}
ds^{2}=-e^{2\phi(r)}\,dt^{2}+\frac{dr^{2}}{1-b(r)/r}+r^{2}\,(d\theta^{2}%
+\sin^{2}{\theta}\,d\varphi^{2})\,,
\end{equation}
representing a spherically symmetric and static wormhole\cite{MT,Visser}.
$b(r)$ is the shape function, while $\phi\left(  r\right)  $ is the redshift
function. $\phi(r)$ and $b(r)$ are arbitrary functions of the radial
coordinate $r\in\left[  r_{0},+\infty\right)  $. A fundamental property of a
traversable wormhole is that a flaring out condition of the throat, given by
$(b-b^{\prime}r)/b^{2}>0$, must be satisfied\cite{MT,Visser}. Furthermore, at
the throat $b(r_{0})=r_{0}$ and the condition $b^{\prime}(r_{0})<1$ is imposed
to have wormhole solutions. Another condition that needs to be satisfied is
$1-b(r)/r>0$. For the wormhole to be traversable, one must demand that there
are no horizons present, which are identified as the surfaces with $e^{2\phi
}\rightarrow0$, so that $\phi(r)$ must be finite everywhere. The last
condition is satisfied if we adopt a Zero Tidal Forces model (ZTF) represented
by $\phi^{\prime}\left(  r\right)  =0$. Such a condition can be imposed by
means of an inhomogeneous Equation of State (EoS) of the form%
\begin{equation}
p_{r}\left(  r\right)  =\omega\left(  r\right)  \rho\left(  r\right)
\label{Inhom}%
\end{equation}
which implies%
\begin{equation}
b\left(  r\right)  +\kappa p_{r}\left(  r\right)  r^{3}=0,\label{constr}%
\end{equation}
where we have used Eqs.$\left(  \ref{rho}\right)  $ and $\left(
\ref{pr0}\right)  $. Eqs.$\left(  \ref{Inhom}\right)  $ and $\left(
\ref{constr}\right)  $ lead to%
\begin{equation}
\omega\left(  r\right)  =-\frac{b\left(  r\right)  }{b^{\prime}\left(
r\right)  r}.\label{o(r)}%
\end{equation}
In Ref.\cite{EPJC}, we have found that the Casimir wormhole described by%
\begin{equation}
\phi\left(  r\right)  ={\ln\left(  \frac{4r}{3r+r_{0}}\right)  \qquad
}\mathrm{and\qquad}b(r)=\frac{2r_{0}}{3}+\frac{r_{0}^{2}}{3r},\label{CWo}%
\end{equation}
does not satisfy the ZTF condition. In this paper, we will consider the
Casimir wormhole shape function deformed by a Yukawa profile satisfying also
the ZTF condition: in this way we have the possibility of building a new
family of solutions which have a vanishing redshift function. We also assume
that the Casimir relationship $\omega=3$ holds, at least on the throat. There
exists another reason to consider a Yukawa deformation to the Casimir
wormhole. Indeed, in Ref.\cite{MG15} we have considered a shape function of
the form%
\begin{equation}
b(r)=r_{0}\exp\left(  -\mu\left(  r-r_{0}\right)  \right)  \label{b(r)00}%
\end{equation}
obeying Eq.$\left(  \ref{o(r)}\right)  $ with\footnote{See also
Ref.\cite{OGenc} for another derivation of a Yukawa shape funtion like the one
in Eq.$\left(  \ref{b(r)00}\right)  $.}%
\begin{equation}
\omega\left(  r\right)  =\frac{1}{\mu r}\label{o(r)Y}%
\end{equation}
and therefore satisfying the ZTF property. However, if we simply assume that%
\begin{equation}
\mu=r_{0}\kappa\rho_{C},
\end{equation}
where%
\begin{equation}
\rho_{C}=\frac{\hbar c\pi^{2}}{720d^{4}},\label{rhoC}%
\end{equation}
then the energy density on the throat becomes%
\begin{equation}
\rho\left(  r\right)  =-\frac{r_{0}\mu}{\kappa r^{2}}\exp\left(  -\mu\left(
r-r_{0}\right)  \right)  \underset{r=r_{0}}{=}-\frac{\mu}{\kappa r_{0}}%
=-\rho_{C},
\end{equation}
namely the Casimir energy density. Moreover, with the help of the relationship
$\left(  \ref{o(r)Y}\right)  $ on the throat one gets%
\begin{equation}
\omega\left(  r_{0}\right)  =\frac{1}{\mu r_{0}}=\frac{1}{r_{0}^{2}\kappa
\rho_{C}}%
\end{equation}
and by imposing that $\omega\left(  r_{0}\right)  =3$, one finds%
\begin{equation}
r_{0}=\sqrt{\frac{1}{3\kappa\rho_{C}}}=\frac{d^{2}}{l_{P}\pi}\sqrt{\frac
{30}{\pi}},\label{omega}%
\end{equation}
in agreement with what found in Ref.\cite{EPJC1} but with a factor $\sqrt{3}$
missing. This example suggests that the mixing between the Casimir wormhole
and a Yukawa wormhole seems to be promising. The paper is organized as
follows: in section \ref{p1} we study three different combinations of the
Casimir wormhole shape function with a Yukawa term, in section \ref{p2} we
explore the consequences of a superposition of the profiles considered in
section \ref{p1}, in section \ref{p3} we adopt the reverse procedure, namely
we fix the form of the energy density and we deduce the form of the shape
function, investigating three different profiles. We summarize and conclude in
section \ref{p4}. Units in which $\hbar=c=k=1$ are used throughout the paper
and will be reintroduced whenever it is necessary.

\section{Casimir-Yukawa Wormholes}

\label{p1}The Casimir wormhole obtained in Ref.\cite{EPJC} has as a source the
original Casimir energy density with a slight but fundamental difference: the
plates separation has been promoted to be a variable instead of being a fixed
quantity. To satisfy the EFE a non vanishing redshift function has been
computed described in Eq.$\left(  \ref{CWo}\right)  $. In this section we are
interested in examining some modifications of the original Casimir wormhole
shape function satisfying the ZTF condition, which can be obtained with the
help of the EoS $\left(  \ref{Inhom}\right)  $. We will take under
consideration three shape function profiles. We begin with

\subsection{$b\left(  r\right)  =r_{0}\exp\left(  -\mu\left(  r-r_{0}\right)
\right)  \left(  2+r_{0}/3r\right)  $}

\label{p1a}The shape function of the Casimir wormhole is defined by%
\begin{equation}
b\left(  r\right)  =\frac{2r_{0}}{3}+\frac{r_{0}^{2}}{3r}.
\end{equation}
We wonder what are the effects of an additional Yukawa term on the original
Casimir shape function whose profile becomes%
\begin{equation}
b\left(  r\right)  =\left(  \frac{2r_{0}}{3}+\frac{r_{0}^{2}}{3r}\right)
\exp\left(  -\mu\left(  r-r_{0}\right)  \right)  ,\label{b(r)0}%
\end{equation}
where $\mu$ is a positive mass scale to be identified. The original Casimir
shape function can be re-obtained when\textrm{\ }$\mu=0$. The profile $\left(
\ref{b(r)0}\right)  $ satisfies the usual properties, namely the throat
condition $b\left(  r_{0}\right)  =r_{0}$, the asymptotic flatness and the
flare out condition of the throat, written into the form%
\begin{equation}
b^{\prime}\left(  r_{0}\right)  =-\frac{1}{3}\left(  1+3\mu r_{0}\right)  <1.
\end{equation}
This is always satisfied together with the property $1-b(r)/r>0$. Another
additional property is%
\begin{equation}
\qquad b\left(  r\right)  \rightarrow0\qquad\mathrm{when\qquad}\mu
\rightarrow\infty\qquad\mathrm{and\qquad}r\rightarrow\infty.\label{as}%
\end{equation}
The energy density can be easily computed and we obtain%
\begin{align}
\rho\left(  r\right)   &  =\frac{b^{\prime}}{\kappa r^{2}}=-\frac{r_{0}%
}{3\kappa r^{4}}\left(  2\mu r^{2}+\mu rr_{0}+r_{0}\right)  \exp\left(
-\mu\left(  r-r_{0}\right)  \right) \nonumber\\
&  =-\frac{1}{\kappa r^{2}}\left(  \mu b\left(  r\right)  +\frac{r_{0}^{2}%
}{3r^{2}}\exp\left(  -\mu\left(  r-r_{0}\right)  \right)  \right)
.\label{rho0}%
\end{align}
It is straightforward to see that, for $\mu\rightarrow0$, one gets the
original Casimir energy density with the plates separation considered as a
variable if we make the following identification
\begin{equation}
\rho\left(  r\right)  =-\frac{r_{0}^{2}}{3\kappa r^{4}}=-\frac{r_{1}^{2}%
}{\kappa r^{4}}=-\frac{\hbar c\pi^{2}}{720r^{4}},
\end{equation}
which is possible if\cite{EPJC}%
\begin{equation}
r_{0}^{2}=3r_{1}^{2}.\label{CW}%
\end{equation}
However, the identification $\left(  \ref{CW}\right)  $ is inconsistent with
the assumption $\left(  \ref{o(r)}\right)  $ because the relationship $\left(
\ref{o(r)}\right)  $ leads to a vanishing redshift, while the identification
$\left(  \ref{CW}\right)  $ does not, as shown in Ref.\cite{EPJC} Therefore,
we consider the following assumption%
\begin{equation}
\rho\left(  r_{0}\right)  =-\frac{\mu}{\kappa r_{0}}-\frac{1}{3\kappa
r_{0}^{2}}=-\frac{\hbar c\pi^{2}}{720d^{4}},\label{rho(r0)0}%
\end{equation}
where $d$ is the \textquotedblleft\textit{fixed plate distance}%
\textquotedblright. This identification fixes the scale mass $\mu$ to the
following value%
\begin{equation}
\mu=\frac{\kappa r_{0}\hbar c\pi^{2}}{720d^{4}}-\frac{1}{3r_{0}}=\frac
{r_{0}l_{P}^{2}\pi^{3}}{90d^{4}}-\frac{1}{3r_{0}}.
\end{equation}
Since $\mu\geq0$, one finds that%
\begin{equation}
\mu=0\qquad\mathrm{when}\qquad r_{0}=\frac{d^{2}}{l_{P}\pi}\sqrt{\frac{30}%
{\pi}},\label{mu0}%
\end{equation}
which is in agreement with what found in Ref.\cite{EPJC1} but with a factor
$\sqrt{3}$ missing. Note that%
\begin{equation}
\lim_{r\rightarrow r_{0}}\lim_{\mu\rightarrow\infty}\rho\left(  r\right)
\neq\lim_{\mu\rightarrow\infty}\lim_{r\rightarrow r_{0}}\rho\left(  r\right)
,\label{limlim}%
\end{equation}
while%
\begin{equation}
\lim_{r\rightarrow r_{0}}\lim_{\mu\rightarrow0}\rho\left(  r\right)
=\lim_{\mu\rightarrow0}\lim_{r\rightarrow r_{0}}\rho\left(  r\right)  .
\end{equation}
Note also that%
\begin{equation}
\lim_{\mu\rightarrow\infty}\rho\left(  r\right)  =0.
\end{equation}
However, due to the relationship $\left(  \ref{mu0}\right)  $, $\mu
\rightarrow\infty$ is equivalent to $r_{0}\rightarrow\infty$. Therefore this
limiting value will be discarded. The second EFE $\left(  \ref{pr0}\right)  $
determines the value of the pressure that, differently from the Casimir
wormhole, will be computed by imposing the relationship $\left(
\ref{o(r)}\right)  $. A simple calculation gives%
\begin{equation}
\omega\left(  r\right)  =\frac{2r+r_{0}}{2\mu r^{2}+\mu rr_{0}+r_{0}%
}.\label{o(r)0}%
\end{equation}
$\omega\left(  r\right)  $ has the following properties%
\begin{gather}
\omega\left(  r_{0}\right)  =\frac{3}{3\mu r_{0}+1}\label{o(r)0b}\\
\omega\left(  r\right)  \underset{r\rightarrow\infty}{\longrightarrow
}0\label{o(r)0c}\\
\omega\left(  r\right)  \underset{\mu\rightarrow0}{\longrightarrow}%
\frac{2r+r_{0}}{r_{0}}\label{o(r)0d}\\
\lim_{r\rightarrow r_{0}}\lim_{\mu\rightarrow0}\omega\left(  r\right)
=\lim_{\mu\rightarrow0}\lim_{r\rightarrow r_{0}}\omega\left(  r\right)
=3,\label{o(r)0e}%
\end{gather}
which is the original relationship between the energy density and the
pressure: in this case the radial pressure. With this assumption, we get%
\begin{equation}
p_{r}(r)=-\frac{1}{\kappa r^{3}}\left(  \frac{2r_{0}}{3}+\frac{r_{0}^{2}}%
{3r}\right)  \exp\left(  -\mu\left(  r-r_{0}\right)  \right)
\end{equation}
and when%
\begin{equation}
\mu\rightarrow0,\qquad p_{r}(r)=-\frac{1}{\kappa r^{3}}\left(  \frac{2r_{0}%
}{3}+\frac{r_{0}^{2}}{3r}\right)  \underset{r\rightarrow r_{0}}{=}-\frac
{1}{\kappa r_{0}^{2}},
\end{equation}
It remains to compute the transverse pressure%
\begin{gather}
p_{t}(r)=\frac{b(r)-b^{\prime}(r)r}{2\kappa r^{3}}=\frac{r_{0}}{6\kappa r^{4}%
}\left(  2\mu r^{2}+\mu rr_{0}+2r+2r_{0}\right)  \exp\left(  -\mu\left(
r-r_{0}\right)  \right) \nonumber\\
=\frac{1}{2\kappa r^{2}}\left(  b(r)\left(  \mu+\frac{1}{r}\right)
+\frac{r_{0}^{2}}{3r^{2}}\exp\left(  -\mu\left(  r-r_{0}\right)  \right)
\right)  ,
\end{gather}
which has the following features, for%
\begin{equation}
\mu\rightarrow0,\qquad p_{t}(r)=\frac{r_{0}}{3\kappa r^{4}}\left(
r+r_{0}\right)  \underset{r\rightarrow r_{0}}{=}\frac{2}{3\kappa r_{0}^{2}},
\end{equation}
The SET becomes%
\begin{equation}
T_{\mu\nu}=T_{\mu\nu}^{a}+T_{\mu\nu}^{b}%
\end{equation}
where%
\begin{equation}
T_{\mu\nu}^{a}=\frac{b\left(  r\right)  }{\kappa r^{2}}\left[  diag\left(
-\mu,-\frac{1}{r},\frac{1}{2}\left(  \mu+\frac{1}{r}\right)  ,\frac{1}%
{2}\left(  \mu+\frac{1}{r}\right)  \right)  \right]
\end{equation}
and%
\begin{equation}
T_{\mu\nu}^{b}=\frac{1}{\kappa r^{2}}\left[  diag\left(  -1,0,\frac{1}{2\kappa
r^{2}},\frac{1}{2\kappa r^{2}}\right)  \right]  \frac{r_{0}^{2}}{3r^{2}}%
\exp\left(  -\mu\left(  r-r_{0}\right)  \right)
\end{equation}
On the throat the SET reduces to%
\begin{equation}
T_{\mu\nu}=\frac{1}{\kappa r_{0}}\left[  diag\left(  -\mu-\frac{1}{3r_{0}%
},-\frac{1}{r_{0}},\frac{1}{2}\left(  \mu+\frac{4}{3r_{0}}\right)  ,\frac
{1}{2}\left(  \mu+\frac{4}{3r_{0}}\right)  \right)  \right]
\end{equation}
and in the limit $\mu\rightarrow0$, one gets%
\begin{equation}
T_{\mu\nu}=\frac{1}{3\kappa r_{0}^{2}}\left[  diag\left(  -1,-3,2,2\right)
\right]  =\frac{\hbar c\pi^{2}}{720d^{4}}\left[  diag\left(  -1,-3,2,2\right)
\right] \label{Tmn}%
\end{equation}
which is verified when the relationship $\left(  \ref{mu0}\right)  $ is
satisfied. Moreover the SET $\left(  \ref{Tmn}\right)  $ is in agreement with
the SET structure found in Ref.\cite{EPJC}. It is interesting to note that,
for $r\rightarrow\infty$, the SET vanishes reproducing a Minkowski SET. With
an abuse of language, one can say that in this limit we find a behavior that
looks like a Generalized Absurdly Benign Traversable Wormhole\cite{EPJC1}. We
say \textquotedblleft\textit{it looks like}\textquotedblright\ because the SET
vanishes for a limiting value of the radial coordinate and not for a well
determined location in space time. The next profile we are going to examine is

\subsection{$b\left(  r\right)  =r_{0}\left(  2\exp\left(  -\mu\left(
r-r_{0}\right)  \right)  +r_{0}/r\right)  /3$}

\label{p1b}For the following profile%
\begin{equation}
b\left(  r\right)  =\frac{2r_{0}}{3}\exp\left(  -\mu\left(  r-r_{0}\right)
\right)  +\frac{r_{0}^{2}}{3r},\label{b(r)1}%
\end{equation}
the Yukawa modification is not distributed over the whole original shape
function but only on the constant term. This little displacement has an
interesting consequence, because when $\mu\rightarrow\infty$ we obtain the
Ellis-Bronnikov (EB)-like wormhole\cite{ellisGL,Bronnikov}. Indeed, the EB
wormhole is%
\begin{equation}
b\left(  r\right)  =\frac{r_{0}^{2}}{r}.\label{EB}%
\end{equation}
The shape function $\left(  \ref{b(r)1}\right)  $ satisfies the usual
properties, namely the throat condition and so on. For completeness, we write
the expression of the flare-out condition, which is%
\begin{equation}
b^{\prime}\left(  r\right)  =-\frac{2r_{0}\mu}{3}\exp\left(  -\mu\left(
r-r_{0}\right)  \right)  -\frac{r_{0}^{2}}{3r^{2}}\underset{r=r_{0}}{=}%
-\frac{2r_{0}\mu+1}{3}<1.
\end{equation}
Even in this case, we can easily compute the energy density to obtain%
\begin{gather}
\rho\left(  r\right)  =\frac{b^{\prime}}{\kappa r^{2}}=\frac{1}{\kappa r^{2}%
}\left(  -\frac{2\mu r_{0}}{3}\exp\left(  -\mu\left(  r-r_{0}\right)  \right)
-\frac{r_{0}^{2}}{3r^{2}}\right) \nonumber\\
=-\frac{1}{\kappa r^{2}}\left(  \mu b\left(  r\right)  +\frac{r_{0}^{2}%
}{3r^{2}}\left(  1-\mu r\right)  \right)
\end{gather}
which, on the throat becomes
\begin{equation}
\rho\left(  r_{0}\right)  =-\frac{\mu2r_{0}+1}{3\kappa r_{0}^{2}}.
\end{equation}
To fix the value of $\mu$ we adopt the same procedure of subsection \ref{p1a}
and we find that even in this case the relationship $\left(  \ref{mu0}\right)
$ is satisfied. The pressure can be determined by solving the second EFE
$\left(  \ref{pr0}\right)  $ and by imposing that the relationship $\left(
\ref{o(r)}\right)  $ be satisfied, namely%
\begin{equation}
\omega\left(  r\right)  =\frac{2r\exp\left(  -\mu\left(  r-r_{0}\right)
\right)  +r_{0}}{2\mu r^{2}\exp\left(  -\mu\left(  r-r_{0}\right)  \right)
+r_{0}}.\label{o(r)1}%
\end{equation}
This time $\omega\left(  r\right)  $ has the following properties%
\begin{gather}
\omega\left(  r_{0}\right)  =\frac{3}{2\mu r_{0}+1}\label{o(r)1a}\\
\omega\left(  r\right)  \underset{\mu\rightarrow\infty}{\longrightarrow
}=1\label{o(r)1b}\\
\omega\left(  r\right)  \underset{r\rightarrow\infty}{\longrightarrow
}=1\label{o(r)1c}\\
\omega\left(  r\right)  \underset{\mu\rightarrow0}{\longrightarrow}%
=\frac{2r+r_{0}}{r_{0}}\label{o(r)1d}\\
\lim_{r\rightarrow r_{0}}\lim_{\mu\rightarrow0}\omega\left(  r\right)
=\lim_{\mu\rightarrow0}\lim_{r\rightarrow r_{0}}\omega\left(  r\right)
=3,\label{o(r)1e}%
\end{gather}
and even in this case the original relationship between the energy density and
the pressure is preserved. Thus the radial pressure is%
\begin{equation}
p_{r}(r)=-\frac{1}{\kappa r^{3}}\left(  \frac{2r_{0}}{3}\exp\left(
-\mu\left(  r-r_{0}\right)  \right)  +\frac{r_{0}^{2}}{3r}\right)
\end{equation}
and one finds that%
\begin{equation}
p_{r}(r)\underset{\mu\rightarrow0}{=}-\frac{r_{0}}{3\kappa r^{3}}\left(
2+\frac{r_{0}}{r}\right)  .
\end{equation}
The last quantity to compute is $p_{t}(r)$, namely%
\begin{equation}
p_{t}(r)=\frac{b(r)-b^{\prime}(r)r}{2\kappa r^{3}}=\frac{3rb(r)\left(  1+\mu
r\right)  +r_{0}^{2}\left(  1-\mu r\right)  }{6\kappa r^{4}},
\end{equation}
which has the following features, for%
\begin{equation}
\mu\rightarrow0,\qquad p_{t}(r)=\frac{r_{0}}{3\kappa r^{4}}\left(
r+r_{0}\right)  \underset{r\rightarrow r_{0}}{=}\frac{2}{3\kappa r_{0}^{2}}.
\end{equation}
To summarize the SET\ for this particular shape function becomes%
\begin{equation}
T_{\mu\nu}=T_{\mu\nu}^{a}+T_{\mu\nu}^{b}%
\end{equation}
where%
\begin{equation}
T_{\mu\nu}^{a}=\frac{b\left(  r\right)  }{\kappa r^{2}}\left[  diag\left(
-\mu,-\frac{1}{r},\frac{1}{2r}\left(  1+\mu r\right)  ,\frac{1}{2r}\left(
1+\mu r\right)  \right)  \right]
\end{equation}
and%
\begin{equation}
T_{\mu\nu}^{b}=\frac{1}{\kappa r^{2}}\left[  diag\left(  -1,0,\frac{1}%
{2},\frac{1}{2}\right)  \right]  \frac{r_{0}^{2}}{3r^{2}}\left(  1-\mu
r\right)
\end{equation}
On the throat the SET reduces to%
\begin{equation}
T_{\mu\nu}=\frac{1}{3\kappa r_{0}^{2}}\left[  diag\left(  -2\mu r_{0}%
-1,-3,2+\mu r_{0},2+\mu r_{0}\right)  \right]
\end{equation}
and in the limit $\mu\rightarrow0$, one gets%
\begin{equation}
T_{\mu\nu}=\frac{1}{3\kappa r_{0}^{2}}\left[  diag\left(  -1,-3,2,2\right)
\right]  =\frac{\hbar c\pi^{2}}{720d^{4}}\left[  diag\left(  -1,-3,2,2\right)
\right]
\end{equation}
which is in agreement with the SET structure found in Ref.\cite{EPJC} only for
$\mu=0$. Finally, we investigate the following shape function

\subsection{$b\left(  r\right)  =r_{0}\left(  2+r_{0}\exp\left(  -\mu\left(
r-r_{0}\right)  \right)  /r\right)  /3$}

\label{p1c}For the following profile%
\begin{equation}
b\left(  r\right)  =\frac{2r_{0}}{3}+\frac{r_{0}^{2}}{3r}\exp\left(
-\mu\left(  r-r_{0}\right)  \right)  ,
\end{equation}
the Yukawa modification is now put only on the variable term. Even in this
modification, we have an interesting consequence, because when $\mu
\rightarrow\infty$ we obtain a constant term smaller than the throat. The
shape function $\left(  \ref{b(r)1}\right)  $ satisfies the usual properties,
namely the throat condition and so on. For completeness, we verify if the
flare out condition is satisfied. We find that%
\begin{gather}
b^{\prime}\left(  r\right)  =-\frac{r_{0}^{2}}{3r^{2}}\left(  1+\mu r\right)
\exp\left(  -\mu\left(  r-r_{0}\right)  \right) \nonumber\\
=\left(  \frac{2}{3}r_{0}-b\left(  r\right)  \right)  \frac{1+\mu r}{r},
\end{gather}
and on the throat one gets%
\begin{equation}
b^{\prime}\left(  r_{0}\right)  =-\frac{1}{3}\left(  1+\mu r_{0}\right)  <1.
\end{equation}
The energy density is straightforward to obtain since%
\begin{gather}
\rho\left(  r\right)  =\frac{b^{\prime}}{\kappa r^{2}}=-\frac{r_{0}^{2}%
}{3\kappa r^{4}}\left(  \mu r+1\right)  \exp\left(  -\mu\left(  r-r_{0}%
\right)  \right) \nonumber\\
=\frac{\mu r+1}{\kappa r^{3}}\left(  \frac{2}{3}r_{0}-b\left(  r\right)
\right) \label{rho2}%
\end{gather}
and for $\mu\rightarrow\infty$, one finds%
\begin{equation}
\rho\left(  r\right)  =0.
\end{equation}
On the throat we obtain%
\begin{equation}
\rho\left(  r_{0}\right)  =-\frac{\mu r_{0}+1}{3\kappa r_{0}^{2}}%
\end{equation}
The second Einstein's field equation $\left(  \ref{pr0}\right)  $ determines
the value of the pressure that and, even in this case, we impose that the
relationship $\left(  \ref{o(r)}\right)  $ be satisfied. This implies that the
redshift function vanishes and that%
\begin{equation}
\omega\left(  r\right)  =\frac{2r\exp\left(  \mu\left(  r-r_{0}\right)
\right)  +r_{0}}{\left(  \mu r+1\right)  r_{0}}.\label{o(r)2}%
\end{equation}
This time $\omega\left(  r\right)  $ has the following properties%
\begin{gather}
\omega\left(  r_{0}\right)  =\frac{3}{\mu r_{0}+1}\label{o(r)2a}\\
\omega\left(  r\right)  \underset{\mu\rightarrow\infty}{\longrightarrow
}=\infty\label{o(r)2b}\\
\omega\left(  r\right)  \underset{r\rightarrow\infty}{\longrightarrow}%
=\infty\label{o(r)2c}\\
\omega\left(  r\right)  \underset{\mu\rightarrow0}{\longrightarrow}%
=\frac{2r+r_{0}}{r_{0}}\label{o(r)2d}\\
\lim_{r\rightarrow r_{0}}\lim_{\mu\rightarrow0}\omega\left(  r\right)
=\lim_{\mu\rightarrow0}\lim_{r\rightarrow r_{0}}\omega\left(  r\right)
=3.\label{o(r)2e}%
\end{gather}
As we can see, from the relationship $\left(  \ref{o(r)2c}\right)  $, one
finds that $\omega\left(  r\right)  $ is divergent: this is a consequence of
the EoS. Indeed for $r\rightarrow\infty$, the energy density $\left(
\ref{rho2}\right)  $ vanishes because of the presence of the damping
exponential overall, while into the pressure the damping exponential appears
only in the constant term. For this reason, this profile will be discarded. In
the next section, we explore a profile which is a superposition of the
previous profiles with the aim of generalizing as much as possible the
features of a Yukawa-Casimir wormhole.

\section{Superposing Traversable Wormholes shape functions}

\label{p2}In this section we will consider a linear combination of the
previous profiles described by the following shape function%
\begin{equation}
b(r)=r_{0}\left(  \alpha\exp\left(  -\mu\left(  r-r_{0}\right)  \right)
+\left(  1-\alpha\right)  \left(  \frac{r_{0}}{r}\right)  ^{c}\exp\left(
-\nu\left(  r-r_{0}\right)  \right)  \right)  ,\label{Yb(r)0}%
\end{equation}
with $\mu,\nu>0,\alpha\geq0$ and $c\in%
\mathbb{R}
$. Note that when $\alpha=2/3$, $c=1$ and $\mu=\nu=0$, it is immediate to see
that the Casimir wormhole shape function is obtained. When $\alpha=1$, we find
a pure Yukawa wormhole discussed in Ref.\cite{MG15} as well as for $c=0$ and
$\mu=\nu$. For $\alpha=0$ and $c=1$, one finds the Yukawa modification to the
EB wormhole. Finally, note that for $\mu=\nu=0$ and $c<-1$, the wormhole is no
more traversable. As a first step we examine under what conditions the
flare-out property is satisfied. From%
\begin{equation}
b^{\prime}(r)=-r_{0}\mu\alpha\exp\left(  -\mu\left(  r-r_{0}\right)  \right)
-\left(  1-\alpha\right)  \left(  \frac{r_{0}}{r}\right)  ^{c}\exp\left(
-\nu\left(  r-r_{0}\right)  \right)  \left(  \nu r_{0}+c\frac{r_{0}}%
{r}\right)  ,\label{Sb'}%
\end{equation}
one finds%
\begin{equation}
b^{\prime}(r_{0})<1\qquad\Longleftrightarrow\qquad\alpha\left(  c+r_{0}%
\nu\right)  <1+r_{0}\left(  \alpha\mu+\nu\right)  +c.
\end{equation}
From the equation $\left(  \ref{Sb'}\right)  $ we can easily compute the
energy density
\begin{equation}
\rho\left(  r\right)  =\frac{b^{\prime}}{\kappa r^{2}}=-\frac{1}{\kappa r^{2}%
}\left[  r_{0}\mu\alpha\exp\left(  -\mu\left(  r-r_{0}\right)  \right)
+\left(  1-\alpha\right)  \left(  \frac{r_{0}}{r}\right)  ^{c}\exp\left(
-\nu\left(  r-r_{0}\right)  \right)  \left(  \nu r_{0}+c\frac{r_{0}}%
{r}\right)  \right]
\end{equation}
and on the throat we obtain%
\begin{equation}
\rho\left(  r_{0}\right)  =-\frac{1}{\kappa r_{0}^{2}}\left[  r_{0}\mu
\alpha-\left(  1-\alpha\right)  \left(  \nu r_{0}+c\right)  \right]
.\label{rhoS}%
\end{equation}
As we can see, $\rho\left(  r_{0}\right)  $ can be considered as a function of
the throat. In order to fix the wormhole throat, we find the stationary point
of $\rho\left(  r_{0}\right)  $, assuming%
\begin{equation}
\rho^{\prime}\left(  r_{0}\right)  =0\qquad\Longrightarrow\qquad\bar{r}%
_{0}=\frac{2c\left(  \alpha-1\right)  }{\mu\alpha-\left(  \alpha-1\right)
\nu}.\label{sta}%
\end{equation}
Plugging $\bar{r}_{0}$ into Eq.$\left(  \ref{rhoS}\right)  $, one finds%
\begin{equation}
\rho\left(  \bar{r}_{0}\right)  =-\frac{\left(  \left(  \mu-\nu\right)
\alpha+\nu\right)  ^{2}}{4c\left(  \alpha-1\right)  \kappa}=-\frac{\hbar
c\pi^{2}}{720d^{4}},
\end{equation}
where we have imposed that, even in this case, the source is described by
Eq.$\left(  \ref{rhoC}\right)  $. A solution of the previous equation is given
by%
\begin{equation}
\bar{\mu}=\frac{\nu}{\alpha}\left(  \alpha-1\right)  \pm\frac{\pi l_{p}%
}{3\alpha d^{2}}\sqrt{c\left(  \alpha-1\right)  \frac{2\pi}{5}},\qquad%
\begin{array}
[c]{c}%
c>0,\qquad\alpha>1\\
c<0,\qquad\alpha<1
\end{array}
.\label{mu}%
\end{equation}
Plugging $\bar{\mu}$ into $\bar{r}_{0}$ of Eq.$\left(  \ref{sta}\right)  $,
one finds%
\begin{equation}
\bar{r}_{0}=\pm\frac{3d^{2}\sqrt{10\pi c\left(  \alpha-1\right)  }}{\pi
^{2}l_{p}}.\label{r0bar}%
\end{equation}
The value of $\alpha$ can be determined with the help of the relationship
$\left(  \ref{o(r)}\right)  $ which, in this case, becomes
\begin{equation}
\omega\left(  r\right)  =\frac{r_{0}\left(  \alpha\exp\left(  -\mu\left(
r-r_{0}\right)  \right)  +\left(  1-\alpha\right)  \left(  \frac{r_{0}}%
{r}\right)  ^{c}\exp\left(  -\nu\left(  r-r_{0}\right)  \right)  \right)
}{r\left[  r_{0}\mu\alpha\exp\left(  -\mu\left(  r-r_{0}\right)  \right)
+\left(  1-\alpha\right)  \left(  \frac{r_{0}}{r}\right)  ^{c}\exp\left(
-\nu\left(  r-r_{0}\right)  \right)  \left(  \nu r_{0}+c\frac{r_{0}}%
{r}\right)  \right]  }\label{o(r)Sp}%
\end{equation}
and on the throat one finds%
\begin{equation}
\omega\left(  r_{0}\right)  =\frac{1}{\left[  r_{0}\mu\alpha+\left(
1-\alpha\right)  \left(  \nu r_{0}+c\right)  \right]  }.
\end{equation}
$\omega\left(  r_{0}\right)  $ can be further reduced to the following simple
expression%
\begin{equation}
\omega\left(  r_{0}\right)  =\frac{1}{\left(  \alpha-1\right)  c}%
=3,\label{o(r0)3}%
\end{equation}
where we have used Eqs. $\left(  \ref{mu}\right)  $ and $\left(
\ref{r0bar}\right)  $ and where we have imposed the Casimir relationship
between pressure and energy density on the throat. Plugging $\left(
\ref{o(r0)3}\right)  $ into $\left(  \ref{r0bar}\right)  $, one gets%
\begin{equation}
\bar{r}_{0}=\frac{\sqrt{30\pi}d^{2}}{\pi^{2}l_{p}}\label{r0Sp}%
\end{equation}
which is the same result of Eq.$\left(  \ref{mu0}\right)  $. With the help of
the Eqs.$\left(  \ref{mu}\right)  $, $\left(  \ref{r0bar}\right)  $ and
$\left(  \ref{o(r0)3}\right)  $, it is possible to show that $\bar{r}_{0}$
represents the minimum of $\rho\left(  r_{0}\right)  $. This means that if we
want to have a wormhole throat with a radius smaller than $\bar{r}_{0}$, we
need to have an increasing of negative energy. To complete the analysis, we
compute the transverse pressure%
\begin{equation}
p_{t}(r)=-{\frac{r_{0}\left(  r_{0}^{c}\left(  \alpha-1\right)  \left(  \nu
{r}^{-c+1}+{r}^{-c}\left(  c+1\right)  \right)  e{^{-\nu\left(  r-r_{0}%
\right)  }}-\alpha e{^{-\mu\left(  r-r_{0}\right)  }}\left(  \mu r+1\right)
\right)  }{2\kappa{r}^{3}}}%
\end{equation}
which, on the throat, becomes%
\begin{equation}
p_{t}(r_{0})={\frac{r_{0}\left(  \left(  \mu-\nu\right)  \alpha+\nu\right)
-c\alpha+c+1}{2\kappa{r}_{0}^{2}}}.
\end{equation}
With the help of the relationships $\left(  \ref{r0Sp}\right)  $, $\left(
\ref{mu}\right)  $ and $\left(  \ref{o(r0)3}\right)  $, one finds%
\begin{equation}
p_{t}(\bar{r}_{0})={\frac{\pi^{3}l_{p}^{2}}{45\kappa d^{4}}=}2\frac{\pi
^{2}\hbar c}{720d^{4}}.
\end{equation}
The analytic form of the SET\ is quite complicated. However, it becomes very
simple on the throat%
\begin{gather}
T_{\mu\nu}^{a}=-\frac{1}{\kappa r_{0}^{2}}\times\nonumber\\
diag\left(  r_{0}\mu\alpha-\left(  1-\alpha\right)  \left(  \nu r_{0}%
+c\right)  ,1,{\frac{r_{0}\left(  \left(  \mu-\nu\right)  \alpha+\nu\right)
-c\alpha+c+1}{2}},{\frac{r_{0}\left(  \left(  \mu-\nu\right)  \alpha
+\nu\right)  -c\alpha+c+1}{2}}\right)
\end{gather}
and in particular in correspondence of the minimum $\bar{r}_{0}$, it
simplifies to%
\begin{equation}
T_{\mu\nu}=\frac{l_{p}^{2}\pi^{3}}{90\kappa d^{4}}\left[  diag\left(
-1,-3,2,2\right)  \right]  =\frac{\hbar c\pi^{2}}{720d^{4}}\left[  diag\left(
-1,-3,2,2\right)  \right]  .
\end{equation}
\textbf{Remark} It is important to observe that thanks to the EoS $\left(
\ref{Inhom}\right)  $, the form of the SET is%
\begin{gather}
T_{\mu\nu}=\frac{r_{0}}{\kappa r^{3}}diag\left(  -\frac{1}{\omega\left(
r\right)  },-1,\frac{1}{2\omega\left(  r\right)  }+\frac{1}{2},\frac
{1}{2\omega\left(  r\right)  }+\frac{1}{2}\right)  \exp\left[  -\int_{r_{0}%
}^{r}\,\frac{d\bar{r}}{\omega(\bar{r})\bar{r}}\right] \nonumber\\
=-\frac{b(r)}{\kappa r^{3}\omega\left(  r\right)  }diag\left(  1,\omega\left(
r\right)  ,-\frac{1}{2}-\frac{\omega\left(  r\right)  }{2},-\frac{1}{2}%
-\frac{\omega\left(  r\right)  }{2}\right) \nonumber\\
=\rho\left(  r\right)  diag\left(  1,\omega\left(  r\right)  ,-\frac{1}%
{2}-\frac{\omega\left(  r\right)  }{2},-\frac{1}{2}-\frac{\omega\left(
r\right)  }{2}\right)  .\label{SET}%
\end{gather}
By construction the SET $\left(  \ref{SET}\right)  $ is divergenceless, but it
is not traceless. However, it is always possible to rearrange the previous SET
$\left(  \ref{SET}\right)  $ in such a way to extract the traceless part.
Indeed%
\begin{equation}
T_{\mu\nu}=T_{\mu\nu}^{T}+\frac{T}{4}g_{\mu\nu}=\frac{\rho\left(  r\right)
}{2}\left[  diag\left(  1,2\omega\left(  r\right)  +1,-\omega\left(  r\right)
,-\omega\left(  r\right)  \right)  -g_{\mu\nu}\right]  ,\label{TmnDec}%
\end{equation}
where $T_{\mu\nu}^{T}$ is the traceless part of the SET $\left(
\ref{SET}\right)  $. It is interesting to observe that by imposing the
following condition%
\begin{equation}
\omega\left(  r_{0}\right)  =1,\label{constr1}%
\end{equation}
one finds that%
\begin{equation}
T_{\mu\nu}^{T}=\frac{\rho\left(  r_{0}\right)  }{2}\left[  diag\left(
1,3,-1,-1\right)  \right]  ,\label{TTmn}%
\end{equation}
independently on the form of $\omega\left(  r\right)  $. This means that
either by decomposing the SET like in Eq.$\left(  \ref{TmnDec}\right)  $ and
fixing $\omega\left(  r_{0}\right)  =1$ or by fixing $\omega\left(
r_{0}\right)  =3$ without the decomposition $\left(  \ref{TmnDec}\right)  $,
it is always possible to preserve the fundamental relationship between
pressure and energy density. Note also that, from the point of view of the
wormholes throat size, the choice $\left(  \ref{constr1}\right)  $ or the
choice $\omega\left(  r_{0}\right)  =3$, do not change the size of the throat
size, as it should be. In the next section, we are going to examine the
reverse procedure, namely we fix the form of the energy density and we will
deduce the form of the shape function.

\section{Traversable Wormholes with a Yukawa Energy Density Profile}

\label{p3}In this section we change the strategy and we fix our attention on
some energy density profiles modified by a Yukawa term and with the help of
Eq.$\left(  \ref{rho}\right)  $, we will deduce the form of the shape
function. Three different forms will be examined. We begin with the following profile

\subsection{$\rho\left(  r\right)  =-r_{0}\rho_{C}\frac{e^{-\mu\left(
r-r_{0}\right)  }}{r}$}

\label{p3a}%

\begin{equation}
\rho\left(  r\right)  =-r_{0}\rho_{C}\frac{e^{-\mu\left(  r-r_{0}\right)  }%
}{r};\qquad\rho_{0}>0,\label{ED1}%
\end{equation}
where $\rho_{C}$ has dimensions of an energy density and $\mu$ is a positive
mass scale parameter. We can identify $\rho_{C}$ with the value expressed by
$\left(  \ref{rhoC}\right)  $. Eq.$\left(  \ref{ED1}\right)  $ can be easily
integrated to obtain
\begin{gather}
b\left(  r\right)  =b\left(  r_{0}\right)  +\kappa\int_{r_{0}}^{r}\rho\left(
r^{\prime}\right)  r^{\prime2}dr^{\prime}\nonumber\\
=r_{0}\left(  1-{\frac{\rho_{C}\kappa}{{\mu}^{2}}-\frac{{r_{0}}\rho_{C}\kappa
}{{\mu}}+\frac{e{^{-\mu\left(  r-r_{0}\right)  }}\rho_{C}\kappa\left(
\mu\,r+1\right)  }{{\mu}^{2}}}\right) \label{b(r)rho}%
\end{gather}
where we have used the condition $b\left(  r_{0}\right)  =r_{0}$. It is
immediate to verify that the shape function $\left(  \ref{b(r)rho}\right)  $
satisfies the asymptotic flatness and the flare-out condition. To have ZTF,
Eq.$\left(  \ref{o(r)}\right)  $ must be imposed, that it means%
\begin{equation}
\omega\left(  r\right)  ={\frac{\left(  {{\mu}^{2}-\rho_{C}{\kappa}}\left(
1+{{r_{0}{\mu}}}\right)  \right)  {e{^{\mu\left(  r-r_{0}\right)  }}}+\rho
_{C}\kappa\left(  \mu\,r+1\right)  }{{\mu}^{2}{r^{2}\rho_{C}\kappa}}}.
\end{equation}
As one can see, for $r\rightarrow\infty$, $\omega\left(  r\right)
\rightarrow\infty$. However, one can adopt another strategy to have a finite
$\omega\left(  r\right)  $. Indeed from the shape function $\left(
\ref{b(r)rho}\right)  $, we can find that there exists $\bar{r}>r_{0}$, such
that $b\left(  \bar{r}\right)  =0$, where%
\begin{equation}
\bar{r}=-{\frac{1}{\mu}\left(  \mathrm{W}\left(  {\frac{{\mu}^{2}-\kappa
{\rho_{C}}\left(  \mu r_{0}+1\right)  }{\kappa{\rho_{C}}\exp\left(  \mu
r_{0}+1\right)  }}\right)  +1\right)  ,}\label{rsol}%
\end{equation}
where $\mathrm{W}\left(  x\right)  $ is the Lambert function defined
mathematically as the multivalued inverse of the function $x\exp\left(
x\right)  $,%
\begin{equation}
W(x)\exp W(x)=x.
\end{equation}
If $-1/e<x<0$, there are two real solutions, and thus two real branches of
$\mathrm{W}$ \cite{Lambert}. Inspired by the Absurdly Benign Traversable
Wormhole (ABTW) and its generalization, the Generalized Absurdly Benign
Traversable Wormhole (GABTW) \cite{EPJC1}, we define the shape function
$\left(  \ref{b(r)rho}\right)  $ in such a way that%
\begin{align}
b\left(  r\right)   &  =r_{0}\left(  1-{\frac{\rho_{C}\kappa}{{\mu}^{2}}%
-\frac{{r_{0}}\rho_{C}\kappa}{{\mu}}+\frac{e{^{-\mu\left(  r-r_{0}\right)  }%
}\rho_{C}\kappa\left(  \mu\,r+1\right)  }{{\mu}^{2}}}\right)  \qquad
\mathrm{for\qquad}r<\bar{r}\nonumber\\
b\left(  r\right)   &  =0\qquad\mathrm{for\qquad}r\geq\bar{r},
\end{align}
where $\bar{r}$ has been defined in $\left(  \ref{rsol}\right)  $. As a
consequence also $\omega\left(  r\right)  $ behaves in the same way and
therefore also the radial pressure. Nevertheless the energy density does not
vanish because $\bar{r}$ does not set to zero its value. However $\rho\left(
\bar{r}\right)  $ can be very small and therefore even the transverse
pressure. Therefore outside the region defined by $r\geq\bar{r}$, one obtains
a quasi-Minkowski spacetime. To complete the analysis, we compute the value of
$\omega\left(  r\right)  $ on the throat. We find%
\begin{equation}
\omega\left(  r_{0}\right)  ={\frac{{1}}{{r_{0}^{2}\rho_{C}\kappa}}%
},\label{OED1}%
\end{equation}
in agreement with what found in Ref.\cite{EPJC1} and with Eq.$\left(
\ref{omega}\right)  $.

\subsection{$\rho\left(  r\right)  =-\frac{\rho_{C}}{2}\left(  \alpha+\beta
r_{0}\frac{e^{-\mu\left(  r-r_{0}\right)  }}{r}\right)  $}

\label{p3b}The second energy density profile we are going to consider is
obtained with a small modification of the profile $\left(  \ref{ED1}\right)  $%
\begin{equation}
\rho\left(  r\right)  =-\frac{\rho_{C}}{2}\left(  \alpha+\beta r_{0}%
\frac{e^{-\mu\left(  r-r_{0}\right)  }}{r}\right)  \qquad\alpha,\beta\in%
\mathbb{R}
.\label{ED2}%
\end{equation}
As we can see, this is a linear combination between the original Casimir
profile and the Yukawa profile $\left(  \ref{ED1}\right)  $. Note that for
$\mu=0$, $\alpha=\beta=1$ and $r=r_{0}$, we obtain the pure Casimir energy
density. Note also that this profile is a generalization of the potential
$\left(  \ref{(2)}\right)  $. Differently from the profile $\left(
\ref{ED1}\right)  $, here we can choose $\alpha$ and $\beta$ in such a way to
have%
\begin{equation}
\rho\left(  \bar{r}\right)  =0\qquad\Longrightarrow\qquad\alpha=-\beta
r_{0}\frac{e^{-\mu\left(  \bar{r}-r_{0}\right)  }}{\bar{r}}\qquad\alpha
,\beta\in%
\mathbb{R}
.\label{rhorbar}%
\end{equation}
The motivation for this choice will be clarified below. The shape function can
be easily computed and we find%
\begin{equation}
b\left(  r\right)  =r_{0}+{\frac{\kappa\rho_{C}\left(  -\alpha\,{\mu}^{2}%
{r}^{3}+\alpha r_{0}^{3}{\mu}^{2}-3\beta r_{0}^{2}\mu-3\beta r_{0}\right)
}{6{\mu}^{2}}+\frac{\kappa\rho_{C}\left(  3\beta\mu rr_{0}+3\beta
r_{0}\right)  e{^{-\mu\,\left(  r-r_{0}\right)  }}}{6{\mu}^{2}}}%
.\label{b(r)ab}%
\end{equation}
It is easy to see that for $r\gg r_{0}$%
\begin{equation}
b\left(  r\right)  \simeq r_{0}-{\frac{\kappa\rho_{C}\left(  \alpha{\mu}%
^{2}{r}^{3}-\alpha r_{0}^{3}{\mu}^{2}+3\beta r_{0}^{2}\mu+3\beta r_{0}\right)
}{6{\mu}^{2}}}\simeq-\frac{\kappa\rho_{C}\alpha}{6}{r}^{3}.
\end{equation}
This behavior is due to the constant term in $\left(  \ref{ED2}\right)  $
which is dominant and produces a divergent shape function. However, since
$\beta$ is not fixed, we can impose that $b\left(  \bar{r}\right)  =0$, where
$\bar{r}$ is the same of Eq.$\left(  \ref{rhorbar}\right)  $. Plugging the
value of $\alpha$ found in Eq.$\left(  \ref{rhorbar}\right)  $ into
Eq.$\left(  \ref{b(r)ab}\right)  $, and by imposing that $b\left(  \bar
{r}\right)  =0 $, one finds%
\begin{equation}
\beta={\frac{6\bar{r}{\mu}^{2}}{\rho\,\kappa\left(  e{^{\mu\,\left(
r_{0}-\bar{r}\right)  }}\left(  {\mu}^{2}r_{0}^{3}-{\mu}^{2}\bar{r}^{3}%
-3\mu\bar{r}^{2}-3\bar{r}\right)  +\left(  3\mu r_{0}+3\right)  \bar
{r}\right)  }}\label{beta}%
\end{equation}
and%
\begin{equation}
b\left(  r\right)  ={\frac{\left(  \left(  3\mu r+3\right)  \bar{r}%
\,e{^{-\mu\,\left(  r-r_{0}\right)  }}+e{^{\mu\,\left(  r_{0}-\bar{r}\right)
}}\left(  {\mu}^{2}{r}^{3}-{\mu}^{2}\bar{r}^{3}-3\,\mu\bar{r}^{2}-3\bar
{r}\right)  \right)  r_{0}}{e{^{\mu\,\left(  r_{0}-\bar{r}\right)  }}\left(
{\mu}r_{0}^{3}-{\mu}^{2}\bar{r}^{3}-3\mu\bar{r}^{2}-3\bar{r}\right)  +\left(
3\mu r_{0}+3\right)  \bar{r}}}\label{b(r)abr1}%
\end{equation}
Thus if we assume that for $r>{\bar{r}}$, $b\left(  \bar{r}\right)  =0$, we
get a feature similar to the ABTW. Moreover, to have ZTF, $\omega\left(
r\right)  $ must be%
\begin{gather}
\omega\left(  r\right)  =-{\frac{b\left(  r\right)  }{b^{\prime}\left(
r\right)  r}}\nonumber\\
{=}\frac{r_{0}6{\mu}^{2}+{\kappa\rho_{C}\left(  -\alpha\,{\mu}^{2}{r}%
^{3}+\alpha r_{0}^{3}{\mu}^{2}-3\beta r_{0}^{2}\mu-3\beta r_{0}\right)
+\kappa\rho_{C}\left(  3\beta\mu rr_{0}+3\beta r_{0}\right)  e{^{-\mu\,\left(
r-r_{0}\right)  }}}}{3r{\mu}^{2}\rho_{C}\left(  \alpha+\beta r_{0}%
\frac{e^{-\mu\left(  r-r_{0}\right)  }}{r}\right)  }%
\end{gather}
and, with the help of Eqs.$\left(  \ref{rhorbar}\right)  $ and $\left(
\ref{beta}\right)  $, we get
\begin{equation}
\omega\left(  r\right)  ={{\frac{\left(  -3\mu r-3\right)  \bar{r}%
e{^{-\mu\,\left(  r-r_{0}\right)  }}-e{^{\mu\,\left(  r_{0}-\bar{r}\right)  }%
}\left(  {\mu}^{2}{r}^{3}-{\mu}^{2}\bar{r}^{3}-3\mu\bar{r}^{2}-3\bar
{r}\right)  }{3\left(  re{^{\mu\,\left(  r_{0}-\bar{r}\right)  }}%
-e{^{-\mu\,\left(  r-r_{0}\right)  }}\bar{r}\right)  {\mu}^{2}{r}^{2}}.}}%
\end{equation}
For $r\rightarrow{\bar{r}}$, $\omega\left(  r\right)  $ is an indeterminate
form of the kind $0/0$. However close to the point $r=\bar{r}$, the shape
function can be approximated by%
\begin{equation}
b\left(  r\right)  \underset{r\rightarrow{\bar{r}}}{\simeq}O\left(  \left(
r-\bar{r}\right)  ^{2}\right)  ,
\end{equation}
while $b^{\prime}\left(  r\right)  $ can be approximated by%
\begin{equation}
b^{\prime}\left(  r\right)  \underset{r\rightarrow{\bar{r}}}{\simeq}%
{\frac{3\bar{r}{\mu}^{2}e{^{\mu\,\left(  r_{0}-\bar{r}\right)  }}\left(
\mu\bar{r}+1\right)  r_{0}}{e{^{\mu\,\left(  r_{0}-\bar{r}\right)  }}\left(
{\mu}^{2}r_{0}^{3}-{\mu}^{2}\bar{r}^{3}-3\mu\bar{r}^{2}-3\bar{r}\right)
+\left(  3\mu r_{0}+3\right)  \bar{r}}}\left(  r-\bar{r}\right)  +O\left(
r-\bar{r}\right)  ^{2}.
\end{equation}
Thus, even in this case, we can assume that
\begin{equation}
\omega\left(  r\right)  =0\qquad r\geq{\bar{r}.}%
\end{equation}
On the throat the analytic form of $\omega\left(  r\right)  $ is far to be
simple. Indeed, we find%
\begin{equation}
\omega\left(  r_{0}\right)  ={{\frac{\left(  -3\mu r_{0}-3\right)  \bar
{r}-e{^{\mu\,\left(  r_{0}-\bar{r}\right)  }}\left(  {\mu}^{2}{r}_{0}^{3}%
-{\mu}^{2}\bar{r}^{3}-3\mu\bar{r}^{2}-3\bar{r}\right)  }{3\left(  r_{0}%
e{^{\mu\,\left(  r_{0}-\bar{r}\right)  }}-\bar{r}\right)  {\mu}^{2}{r}_{0}%
^{2}}.}}%
\end{equation}
However, one finds that%
\begin{equation}
\omega\left(  r_{0}\right)  \underset{\mu\rightarrow0}{\simeq}{{\frac
{-2r_{0}^{2}+r_{0}\bar{r}+\bar{r}^{2}}{6{r}_{0}^{2}}\geq0\qquad\mathrm{for}%
\qquad}\bar{r}}\geq r_{0}{,}%
\end{equation}
while%
\begin{equation}
\omega\left(  r_{0}\right)  \underset{\mu\rightarrow\infty}{\simeq}{{\frac
{1}{r_{0}{\mu}}+{{\frac{1}{{\mu}^{2}{r}_{0}^{2}}}}\rightarrow0}.}%
\end{equation}
However when $\bar{r}\gg r_{0}$, we get%
\begin{equation}
\omega\left(  r_{0}\right)  \underset{\bar{r}\gg r_{0}}{\simeq}{{\frac
{1}{r_{0}{\mu}}+{{\frac{1}{{\mu}^{2}{r}_{0}^{2}},}}}}%
\end{equation}
which is finite and positive. Therefore we can conclude that from an energy
density of the form $\left(  \ref{ED2}\right)  $, it is possible to extract
another shape function which generalizes an ABTW. It is important to observe
that such a generalization is realized because of the presence of a repulsive
Yukawa-Casimir profile, otherwise for a choice of the form%
\begin{equation}
\rho\left(  \bar{r}\right)  =-\frac{\rho_{C}}{2}\left(  1+r_{0}\frac
{e^{-\mu\left(  \bar{r}-r_{0}\right)  }}{r}\right)  \simeq-\frac{\rho_{C}}%
{2}\neq0,
\end{equation}
the energy density outside the region $r>\bar{r}$ is not Minkowski. Although
interesting, the profile $\left(  \ref{ED2}\right)  $ has the defect of having
a way to compare the throat radius with the original Casimir source, like in
Eq.$\left(  \ref{OED1}\right)  $.

\subsection{$\rho\left(  r\right)  =\frac{r_{0}\rho_{C}}{r}\left(  \alpha
e^{-\mu\left(  r-r_{0}\right)  }-\left(  1-\alpha\right)  e^{-\nu\left(
r-r_{0}\right)  }\right)  $}

\label{p3c}To this purpose, we fix our attention on an energy density profile
which can reproduce both a Yukawa behavior and an ABTW. This is represented by%
\begin{equation}
\rho\left(  r\right)  =\frac{r_{0}\rho_{C}}{r}\left(  \alpha e^{-\mu\left(
r-r_{0}\right)  }-\left(  1-\alpha\right)  e^{-\nu\left(  r-r_{0}\right)
}\right)  \qquad\mu,\nu>0.\label{rhoab}%
\end{equation}
When $\alpha=0$, we obtain the profile $\left(  \ref{ED1}\right)  $, while
when $\alpha=1$, we obtain its repulsive version. $\forall\alpha\neq0,1$, we
have a linear superposition of the Yukawa-Casimir profile. The combination of
an attractive and repulsive potential is suggested also by the potential
$\left(  \ref{ED2}\right)  $ together with the option $\left(  \ref{rhorbar}%
\right)  $. Note that the option $\left(  \ref{rhorbar}\right)  $ is relevant
only if one wishes to reproduce an ABTW. If such an option is not adopted the
existence of a traversable wormhole is guaranteed the same. For the profile
$\left(  \ref{rhoab}\right)  $, it is immediate to calculate the form of the
shape function
\begin{gather}
b\left(  r\right)  =r_{0}+{\frac{r_{0}\left(  1+\nu r_{0}\right)  \left(
\alpha-1\right)  \rho_{C}\kappa}{{\nu}^{2}}}+{\frac{\alpha r_{0}\kappa\rho
_{C}\left(  1+{{\mu}r_{0}}\right)  }{{\mu}^{2}}}\nonumber\\
-{\frac{r_{0}\kappa\rho_{C}\left(  \nu r+1\right)  \left(  \alpha-1\right)
e{^{-\nu\,\left(  r-r_{0}\right)  }}}{{\nu}^{2}}}-{\frac{\alpha r_{0}%
\kappa\rho_{C}\left(  \mu r+1\right)  e{^{-\mu\,\left(  r-r_{0}\right)  }}%
}{{\mu}^{2}}}\label{b(r)mn}%
\end{gather}
and for $r\rightarrow\infty$, we find%
\begin{equation}
b\left(  r\right)  \underset{r\rightarrow\infty}{\simeq}r_{0}+{\frac
{r_{0}\left(  1+\nu r_{0}\right)  \left(  \alpha-1\right)  \rho_{C}\kappa
}{{\nu}^{2}}}+{\frac{\alpha r_{0}\kappa\rho_{C}\left(  1+{{\mu}r_{0}}\right)
}{{\mu}^{2}}}%
\end{equation}
which can be set to zero if%
\begin{equation}
\alpha={\frac{{\mu}^{2}\left(  \kappa\nu r_{0}\rho_{C}+\rho_{C}\kappa-{\nu
}^{2}\right)  }{\left(  {\mu}^{2}\nu r_{0}+\mu{\nu}^{2}r_{0}+{\mu}^{2}+{\nu
}^{2}\right)  \rho_{C}\kappa}.}\label{alpha}%
\end{equation}
Plugging Eq.$\left(  \ref{alpha}\right)  $ into $\left(  \ref{b(r)mn}\right)
$, one finds%
\begin{equation}
b\left(  r\right)  =r_{0}{\frac{\left(  e{^{-\nu\left(  r-r_{0}\right)  }%
}\left(  \rho\kappa\left(  1+\mu r_{0}\right)  +{\mu}^{2}\right)  \left(  \nu
r+1\right)  -e{^{-\mu\left(  r-r_{0}\right)  }}\left(  \rho\kappa\left(  1+\nu
r_{0}\right)  -{\nu}^{2}\right)  \left(  \mu r+1\right)  \right)  }{\left(
\nu r_{0}+1\right)  {\mu}^{2}+\mu{\nu}^{2}r_{0}+{\nu}^{2}}}\label{b(r)mn1}%
\end{equation}
which is useful to compute the function of the EoS $\omega\left(  r\right)  $%
\begin{equation}
\omega\left(  r\right)  ={\frac{\left(  \rho\kappa\left(  \mu r_{0}+1\right)
+{\mu}^{2}\right)  e{^{-\nu\left(  r-r_{0}\right)  }}\left(  \nu r+1\right)
-\left(  \left(  \nu r_{0}+1\right)  \rho\kappa-{\nu}^{2}\right)  \left(  \mu
r+1\right)  e{^{-\mu\left(  r-r_{0}\right)  }}}{{r}^{2}\left(  {\nu}%
^{2}\left(  \rho\left(  \mu r_{0}+1\right)  \kappa+{\mu}^{2}\right)
e{^{-\nu\left(  r-r_{0}\right)  }}-{\mu}^{2}\left(  \left(  \nu r_{0}%
+1\right)  \rho\kappa+{\nu}^{2}\right)  e{^{-\mu\left(  r-r_{0}\right)  }%
}\right)  }.}%
\end{equation}
This time the function $\omega\left(  r\right)  $ goes to zero for large
values of $r$, while on the throat one gets%
\begin{equation}
\omega\left(  r_{0}\right)  ={\frac{\left(  1+\nu r_{0}\right)  {\mu}%
^{2}+\left(  1+\mu r_{0}\right)  {\nu}^{2}}{2{\nu}^{2}r_{0}^{2}{\mu}^{2}%
+r_{0}^{3}\rho\kappa{\nu}\mu\left(  {\nu}-{\mu}\right)  +r_{0}^{2}\rho
\kappa\left(  {\nu}^{2}-{\mu}^{2}\right)  }.}%
\end{equation}
Even in this case, if we desire to extract information on the throat size, we
need to compare $\omega\left(  r_{0}\right)  $ with a physical source like the
Casimir source. To do this, we assume that%
\begin{equation}
\omega\left(  r_{0}\right)  =1{.}\label{omegamn}%
\end{equation}
To do calculations in practice, it is useful the following setting%
\begin{equation}
\mu=\frac{m}{r_{0}};\qquad\nu=\frac{n}{r_{0}}\qquad\mathrm{and\qquad}%
r_{0}=\frac{x}{\sqrt{\rho\kappa}};\qquad m,n\in%
\mathbb{R}
_{+}\label{resc}%
\end{equation}
and Eq.$\left(  \ref{omegamn}\right)  $ becomes%
\begin{equation}
{\frac{\left(  1+n\right)  m^{2}+\left(  1+m\right)  n^{2}}{\left(
2n^{2}m^{2}+x^{2}nm\left(  n-m\right)  +x^{2}\left(  n^{2}-m^{2}\right)
\right)  }=1,}%
\end{equation}
whose solution is%
\begin{equation}
x={\frac{\sqrt{\left(  2{n}^{2}-n-1\right)  {m}^{2}-{n}^{2}m-{n}^{2}}}%
{\sqrt{\left(  \left(  n+1\right)  m+n\right)  \left(  m-n\right)  }}%
.}\label{r}%
\end{equation}
To constraint $r$ to be very small, we observe that the r.h.s. of Eq. $\left(
\ref{r}\right)  $ vanishes when%
\begin{equation}
m_{\pm}=n\frac{1\pm\sqrt{9{n}^{2}-4n-4}}{2\left(  2{n}^{2}-n-1\right)  }\qquad
n\geq\frac{2}{9}\left(  1+\sqrt{10}\right)  \simeq0.92495.
\end{equation}
$m_{-}$ will be discarded because is the negative root. Note that it is not
necessary to have a vanishing $x$, rather we need an $x$ with a value of the
order of $10^{-10}$ or greater. This is due to the rescaling in $\left(
\ref{resc}\right)  $ setting the size of the wormhole throat to be of the
order of%
\begin{equation}
r_{0}\simeq x\times10^{17}m.
\end{equation}
Therefore we can conclude that with a linear combination of two Yukawa-Casimir
profiles, actually a difference of them, one finds a traversable wormhole with
a throat that can be fine tuned with respect to the original Casimir source.

\section{Conclusions}

\label{p4}In this paper we have taken under examination the modification of
the Casimir wormhole examined in Ref.\cite{EPJC} which uses, as a source, the
negative energy density of the Casimir device. Differently from
Ref.\cite{EPJC}, this time we have imposed the ZTF condition to obtain
different solutions. We have found that the ZTF condition can be imposed only
if we modify the form of the energy density or the form of the shape function.
To this purpose we have considered Yukawa type modifications of the original
profile. The motivation for this choice stands in the attempt to detect
signals of variations of the ordinary gravitational field even for TW and to
obtain the possibility of having the negative energy density more concentrated
in proximity of the throat. We have divided the paper in two parts: the first
part is devoted to the analysis of the modification of the shape function with
a Yukawa term and the second part is devoted to the modification of the
original Casimir energy density with an appropriate Yukawa term. In the first
part we have examined three different profiles having in common the Casimir
wormhole shape function, namely%
\begin{equation}
b\left(  r\right)  =\frac{2}{3}r_{0}+\frac{r_{0}^{2}}{3r}%
\end{equation}
and we have included a Yukawa modification of the type $\exp\left(
-\mu\left(  r-r_{0}\right)  \right)  $ acting on every single term and
globally. Two of the three profiles have shown features compatible with the
throat size estimated in Ref.\cite{EPJC1}, while one of them has developed a
divergent behavior for large values of the radial variable $r$. I\ recall to
the reader that, in this paper, we have examined the Casimir energy density
with the plates separation considered as a parameter and not as a variable.
This choice has led to have a huge throat size instead of a Planckian one like
in Ref.\cite{EPJC}. As a further analysis, we have also considered a
superposition of different categories of TW. Even in this case the resulting
size of the wormholes throat is huge and compatible with the size of a GABTW
described in Ref.\cite{EPJC1}. I recall again to the reader that the huge size
of the wormhole throat has been found by imposing that the inhomogeneous
function of the EoS $\left(  \ref{o(r)}\right)  $ at the throat has a constant
value compatible with the ordinary Casimir relationship $p=3\rho$. In the
second part, we have fixed the form of the energy density and we have deduced
the form of the shape function with the help of the first of the EFE $\left(
\ref{rho}\right)  $. Even in this case, we have analyzed three different
profiles. Since every of these profiles has produced a correction to the size
of the throat at infinity, we have considered the possibility of taken another
generalization for the ABTW. In particular we have found a value of the radial
variable, located at $r=\bar{r}$ where the shape function vanishes and we have
truncated the region outside $r=\bar{r}$. In this way the pressure outside the
region $r=\bar{r}$ vanishes. However, for the profile of subsection \ref{p3a},
the energy density does not vanish for $r\geq\bar{r}$, it is small because of
the exponential but not nought. Therefore the structure of an ABTW or GABTW
cannot be reproduced. On the contrary, for the profiles discussed in
subsection \ref{p3b} and \ref{p3c}, it is possible to reproduce an ABTW in a
generalized form different by the GABTW at the price of introducing a
repulsive potential, namely we have the difference of two Yukawa profiles: one
attractive and one repulsive. I\ recall the reader that an ABTW is defined by
the following shape and redshift functions%
\begin{align}
b(r) &  =r_{0}\left(  1-\left(  \frac{r-r_{0}}{a}\right)  \right)  ^{2}%
,\qquad\Phi(r)=0;\qquad r_{0}\leq r\leq r_{0}+a\nonumber\\
b(r) &  =0,\qquad\Phi(r)=0;\qquad r\geq r_{0}+a.\label{b(r)AB}%
\end{align}
Therefore outside the location $r=r_{0}+a$, the spacetime is Minkowski. The
same happens for a GABTW, where one finds%
\begin{align}
b(r) &  =r_{0}\frac{\left(  1-\mu\left(  r-r_{0}\right)  \right)  ^{\alpha}%
}{\left(  1-\nu\left(  r-r_{0}\right)  \right)  ^{\beta}},\qquad
\Phi(r)=0;\qquad r_{0}\leq r\leq r_{0}+1/\mu\nonumber\\
b(r) &  =0,\qquad\Phi(r)=0;\qquad r\geq r_{0}+1/\mu.\label{GABTWII}%
\end{align}
Of course $1/\mu$ plays the r\^{o}le of $a$ and vice versa. Note that it is
the exponent in Eq.$\left(  \ref{b(r)AB}\right)  $ and in Eq.$\left(
\ref{GABTWII}\right)  $ that plays a key r\^{o}le to determine the Minkowski
structure for $r\geq r_{0}+a$ or $r\geq r_{0}+1/\mu$. Such a property is
completely absent for the profiles discussed in subsection \ref{p3a},
\ref{p3b} and \ref{p3c} and one must build a profile that potentially can
develop such a property. This is the reason why a repulsive Yukawa profile is
necessary to have a vanishing value outside a certain region. Of course this
is related to the attempt to reproduce some of the features of an ABTW. If one
abandons this request, the repulsive potential is not fundamental. However,
our insistence to reproduce the features of an ABTW is justified by the fact
the negative energy density is concentrated in a very small region of the
space and there is no redshift. Coming back to the profile $\left(
\ref{rhoab}\right)  $ in subsection \ref{p3c}, it represents again a
difference of two Yukawa profiles, and behaves \textquotedblright%
\textit{like}\textquotedblright\ a ABTW, because for a sufficiently large
values of $r$, $\rho\left(  r\right)  $, $p_{r}\left(  r\right)  $ and
$p_{t}\left(  r\right)  $ vanish. Nevertheless, because of the exponentials,
it is not necessary that the radial value $r$ needs to be really large.
Another interesting feature of the profile $\left(  \ref{rhoab}\right)  $ is
that, this time, we can fine tune the throat size down to acceptable values,
which is exactly what one needs. To conclude, we have also to point out that
in the context of \textit{Self-Sustained Traversable Wormholes}, namely TW
sustained by their own quantum
fluctuations\cite{RG,RG1,RGFSNL,RGFSNL1,RGFSNL2}, could be interesting to
consider how the Yukawa-Casimir TW behaves in this context.

\section{Acknowledgements}

This work has been supported by the project "Traversable Wormholes: A Road To
Interstellar Exploration," an Interstellar Initiatives Grant award funded by
the Limitless Space Institute.

\appendix{}

\section{Features of the Superposition of Traversable Wormholes}

\label{p5}In this section we are going to explore some of the features of the
profile $\left(  \ref{Yb(r)0}\right)  $ which is quite general to include the
other profiles discussed in section \ref{p1} of this paper. We begin to
examine the proper length which is defined as%
\begin{equation}
b(r)=r_{0}\left(  \alpha\exp\left(  -\mu\left(  r-r_{0}\right)  \right)
+\left(  1-\alpha\right)  \left(  \frac{r_{0}}{r}\right)  ^{c}\exp\left(
-\nu\left(  r-r_{0}\right)  \right)  \right)  ,
\end{equation}%
\begin{equation}
l\left(  r\right)  =\pm\int_{r_{0}}^{r}\frac{dr^{\prime}}{\sqrt{1-\frac
{b\left(  r^{\prime}\right)  }{r^{\prime}}}},
\end{equation}
where the\textquotedblleft$\pm$\textquotedblright\ depends on the wormhole
side we are. In the case of the shape function $\left(  \ref{Yb(r)0}\right)
$, we know that it vanishes exponentially for $r\rightarrow+\infty$. This is
true for $\mu\neq0$ and $\nu\neq0$. For instance for $\mu=\nu=0$ and $c=1$,
one finds that the shape function is represented by the EB wormhole $\left(
\ref{EB}\right)  $, whose proper length is%
\begin{equation}
l\left(  r\right)  =\pm\sqrt{r^{2}-r_{0}^{2}}.
\end{equation}
For this reason, for the other cases it is sufficient to consider what happens
close to the throat. In general, we can write%
\begin{equation}
\frac{\sqrt{r}}{\sqrt{r-b\left(  r\right)  }}\underset{r\rightarrow r_{0}%
}{\simeq}\frac{\sqrt{r}}{\sqrt{1-b^{\prime}\left(  r_{0}\right)  }%
\sqrt{r-r_{0}}}.
\end{equation}
Thus the approximated proper length becomes%
\begin{gather}
l\left(  r\right)  \underset{r\rightarrow r_{0}}{\simeq}\pm\frac{1}%
{\sqrt{1-b^{\prime}\left(  r_{0}\right)  }}\int_{r_{0}}^{r}\frac
{\sqrt{r^{\prime}}dr^{\prime}}{\sqrt{r^{\prime}-r_{0}}}\nonumber\\
=\pm\frac{r_{0}}{\sqrt{1-b^{\prime}\left(  r_{0}\right)  }}\left[  \sqrt
{\frac{r}{r_{0}}}\sqrt{\frac{r}{r_{0}}-1}+{\ln\left(  \sqrt{\frac{r}{r_{0}}%
}+\sqrt{\frac{r}{r_{0}}-1}\right)  }\right]  .\label{l(r)0}%
\end{gather}
A further approximation leads to%
\begin{equation}
l\left(  r\right)  \underset{r\rightarrow r_{0}}{\simeq}\pm\frac{r_{0}%
\sqrt{r/r_{0}}\sqrt{r/r_{0}-1}}{\sqrt{1-b^{\prime}\left(  r_{0}\right)  }%
}\label{l(r)0a}%
\end{equation}
and, in the case of the shape function $\left(  \ref{Yb(r)0}\right)  $,
$\left(  \ref{l(r)0a}\right)  $ assumes the form%
\begin{equation}
l\left(  r\right)  =\pm\;\frac{r_{0}\sqrt{r/r_{0}}\sqrt{r/r_{0}-1}}%
{\sqrt{1+r_{0}\left(  \alpha\mu+\nu\left(  1-\alpha\right)  \right)  +c\left(
1-\alpha\right)  }}.
\end{equation}
The argument of the denominator is positive if and only if the flare-out
condition is satisfied. However, we can use the constraint $\left(
\ref{mu}\right)  $ and $\left(  \ref{o(r0)3}\right)  $ to have a better
estimate. We find%
\begin{equation}
l\left(  r\right)  ={\frac{6\sqrt{r-r_{0}}\sqrt{r_{0}}\sqrt{5}d}{\sqrt{{\pi
}^{3/2}\sqrt{30}l_{p}r_{0}+30\,{d}^{2}}}}%
\end{equation}
or by means of the constraint $\left(  \ref{r0bar}\right)  $, we can write the
proper length only in terms of the plates separation%
\begin{equation}
l\left(  r\right)  ={\frac{\sqrt{3}d\sqrt[4]{10}}{{\pi}^{{\frac{3}{2}}}l_{p}%
}\sqrt{l_{p}{\pi}^{{\frac{3}{2}}}\sqrt{3}r-3{d}^{2}\sqrt{10}}}.\label{l(r)}%
\end{equation}
In a similar way, to compute the embedded surface, we need to evaluate%

\begin{equation}
z\left(  r\right)  =\pm\int_{r_{0}}^{r}\frac{dr^{\prime}}{\sqrt{\frac
{r^{\prime}}{b\left(  r^{\prime}\right)  }-1}},
\end{equation}
which, close to the throat, becomes%
\begin{gather}
z\left(  r\right)  =\pm\int_{r_{0}}^{r}\frac{\sqrt{b\left(  r^{\prime}\right)
}dr^{\prime}}{\sqrt{r^{\prime}-b\left(  r^{\prime}\right)  }}\underset
{r\rightarrow r_{0}}{\simeq}\pm\frac{\sqrt{r_{0}}}{\sqrt{1-b^{\prime}\left(
r_{0}\right)  }}\int_{r_{0}}^{r}\frac{dr^{\prime}}{\sqrt{r^{\prime}-r_{0}}%
}\nonumber\\
=\pm\frac{2\sqrt{r_{0}}\sqrt{r-r_{0}}}{\sqrt{1+r_{0}\left(  \alpha\mu
+\nu\left(  1-\alpha\right)  \right)  +c\left(  1-\alpha\right)  }%
}.\label{z(r)}%
\end{gather}
To further investigate the properties of the Traversable Wormholes described
by the shape function $\left(  \ref{Yb(r)0}\right)  $, we consider the
computation of the total gravitational energy for a wormhole, defined
as\cite{NZCP}%
\begin{equation}
E_{G}\left(  r\right)  =\int_{r_{0}}^{r}\left[  1-\sqrt{\frac{1}{1-b\left(
r^{\prime}\right)  /r^{\prime}}}\right]  \rho\left(  r^{\prime}\right)
dr^{\prime}r^{\prime2}+\frac{r_{0}}{2G}=M-M_{\pm}^{P},\label{Eg}%
\end{equation}
where%
\begin{equation}
M=\int_{r_{0}}^{r}4\pi\rho\left(  r^{\prime}\right)  r^{\prime2}dr^{\prime
}+\frac{r_{0}}{2G}\label{M}%
\end{equation}
is the total mass and%
\begin{equation}
M_{\pm}^{P}=\pm\int_{r_{0}}^{r}\frac{4\pi\rho\left(  r^{\prime}\right)
r^{\prime2}}{\sqrt{1-b\left(  r^{\prime}\right)  /r^{\prime}}}dr^{\prime
}.\label{Mp}%
\end{equation}
is the proper mass. In particular we find for the total mass%
\begin{equation}
Mc^{2}=\int_{r_{0}}^{r}4\pi\rho\left(  r^{\prime}\right)  r^{\prime
2}dr^{\prime}+\frac{r_{0}}{2G}=\frac{4\pi}{\kappa}\int_{r_{0}}^{r}b^{\prime
}\left(  r^{\prime}\right)  dr^{\prime}+\frac{r_{0}}{2G}=\frac{4\pi}{\kappa
}b\left(  r\right)  \underset{r\rightarrow\infty}{\rightarrow}0\label{M1}%
\end{equation}
where we have used the relationship $\left(  \ref{r0Sp}\right)  $ and we have
momentarily reintroduced the speed of light. For $M_{\pm}^{P}$, we can
estimate the value of the integral close to the throat, following what has
been done for the proper length and the embedded surface. We can write%
\begin{align}
M_{\pm}^{P}c^{2}  &  =\pm\int_{r_{0}}^{r}\frac{4\pi\rho\left(  r^{\prime
}\right)  r^{\prime2}}{\sqrt{1-b\left(  r^{\prime}\right)  /r^{\prime}}%
}dr^{\prime}\underset{r\rightarrow r_{0}}{\simeq}\pm\frac{b^{\prime}\left(
r_{0}\right)  }{2G\sqrt{1-b^{\prime}\left(  r_{0}\right)  }}\int_{r_{0}}%
^{r}\frac{dr^{\prime}}{\sqrt{r-r_{0}}}\nonumber\\
&  \simeq\pm\frac{\left(  \alpha-1\right)  \left(  \nu r_{0}+c\right)
-r_{0}\mu\alpha}{G\sqrt{1+r_{0}\left(  \alpha\mu+\nu\left(  1-\alpha\right)
\right)  +c\left(  1-\alpha\right)  }}\sqrt{r-r_{0}},\label{Mp1}%
\end{align}
where the \textquotedblleft$\pm$\textquotedblright\ depends one the wormhole
side we are. Thus the total gravitational energy $\left(  \ref{Eg}\right)  $
becomes%
\begin{equation}
E_{G}\left(  r\right)  \simeq%
\begin{array}
[c]{c}%
\frac{r_{0}}{2G}\qquad r\rightarrow r_{0}\\
\\
0\qquad r\rightarrow\infty
\end{array}
.
\end{equation}
Even for the total energy, this is true for $\alpha\neq0$, $\nu\neq0$ and
$c\neq1$. Indeed for $\alpha=\nu=0$ and $c=1$ we can write the total
gravitational energy of the EB wormhole $\left(  \ref{EB}\right)  $ which
reduces to%
\begin{equation}
E_{G}\left(  r\right)  =\frac{r_{0}}{3G}\left(  1\mp{\frac{\sqrt{3}{\pi}c^{4}%
}{6}}\right)  .
\end{equation}
It is interesting to note that the total energy is concentrated completely on
the throat and at infinity vanishes showing a screening mechanism: in other
words, the \textquotedblleft\textit{imprint at infinity}\textquotedblright%
\ disappears\cite{Visser}. Another important traversability condition is that
the acceleration felt by the traveller should not exceed Earth's gravity
$g_{\oplus}\simeq980$ $cm/s^{2}$. In an orthonormal basis of the traveller's
proper reference frame, we can find%
\begin{equation}
\left\vert \mathbf{a}\right\vert =\left\vert \sqrt{1-\frac{b\left(  r\right)
}{r}}e^{-\Phi(r)}\left(  \gamma e^{\Phi(r)}\right)  ^{\prime}\right\vert
\leq\frac{g_{\oplus}}{c^{2}}%
\end{equation}
and in this case, because $\Phi(r)=0$, the traveller has no acceleration,
which is in agreement with Ref.\cite{MT}. As regards the lateral tidal forces,
we find%
\begin{gather}
\left\vert \frac{\gamma^{2}c^{2}}{2r^{2}}\left[  \frac{v^{2}\left(  r\right)
}{c^{2}}\left(  b^{\prime}\left(  r\right)  -\frac{b\left(  r\right)  }%
{r}\right)  +2r\left(  r-b\left(  r\right)  \right)  \Phi^{\prime}\left(
r\right)  \right]  \right\vert \left\vert \eta\right\vert \nonumber\\
=\left\vert \frac{\gamma^{2}c^{2}}{2r^{3}}\left[  -\frac{v^{2}\left(
r\right)  b\left(  r\right)  }{c^{2}}\left(  \frac{1}{\omega\left(  r\right)
}+1\right)  \right]  \right\vert \left\vert \eta\right\vert \leq g_{\oplus
},\label{LTC}%
\end{gather}
where we have used the relationship $\left(  \ref{o(r)}\right)  $. This is a
constraint about the velocity with which observers traverse the wormhole.
$\eta$ represents the size of the traveller which can be fixed approximately
equal, at the symbolic value of $2$ $m$\cite{MT}. If we assume a constant
speed $v$ and $\gamma\simeq1$, close to the throat, the lateral tidal
constraint becomes%
\begin{gather}
\left\vert \frac{\gamma^{2}c^{2}}{2r_{0}^{2}}\left[  -\frac{v^{2}\left(
r_{0}\right)  }{c^{2}}\left(  1+\left[  r_{0}\mu\alpha+\left(  1-\alpha
\right)  \left(  \nu r_{0}+c\right)  \right]  \right)  \right]  \right\vert
\left\vert 2\right\vert \nonumber\\
\simeq\left\vert \left[  \frac{v^{2}\left(  r_{0}\right)  }{r_{0}^{2}}\right]
\left(  1+\left[  r_{0}\mu\alpha+\left(  1-\alpha\right)  \left(  \nu
r_{0}+c\right)  \right]  \right)  \right\vert \lesssim g_{\oplus}\nonumber\\
\qquad\Longrightarrow\qquad v\lesssim r_{0}\sqrt{\left(  1+\left[  r_{0}%
\mu\alpha+\left(  1-\alpha\right)  \left(  \nu r_{0}+c\right)  \right]
\right)  g_{\oplus}}.\label{LTCt}%
\end{gather}
If the observer has a vanishing $v$, then the tidal forces are null. Note that
the total time defined by%
\begin{equation}
\Delta t=\int_{r_{0}}^{r}\frac{e^{-\phi\left(  r^{\prime}\right)  }dr^{\prime
}}{v\sqrt{1-\frac{b\left(  r^{\prime}\right)  }{r^{\prime}}}}%
\end{equation}
and the total proper time given by%
\begin{equation}
\Delta\tau=\int_{r_{0}}^{r}\frac{dr^{\prime}}{v\sqrt{1-\frac{b\left(
r^{\prime}\right)  }{r^{\prime}}}}%
\end{equation}
are the same for the profile $\left(  \ref{Yb(r)0}\right)  $ because the
redshift function is nought. Assuming that $v$ is approximately constant, we
can use the estimate $\left(  \ref{LTCt}\right)  $ to complete the evaluation
of the crossing time which approximately is%
\begin{equation}
\Delta\tau=\Delta t\simeq\frac{2\times10^{4}}{\left(  1+r_{0}\left(  \alpha
\mu+\nu\left(  1-\alpha\right)  \right)  +c\left(  1-\alpha\right)  \right)
\sqrt{g_{\oplus}}}.
\end{equation}
However, we can use the constraint $\left(  \ref{sta}\right)  $, $\left(
\ref{mu}\right)  $, $\left(  \ref{r0bar}\right)  $ and $\left(  \ref{o(r0)3}%
\right)  $ to have a better estimate of the crossing time which becomes%
\begin{equation}
\Delta\tau=\Delta t\simeq\frac{3\times10^{4}}{2\sqrt{g_{\oplus}}}%
\simeq4.79\times10^{3}s,
\end{equation}
where we have considered a possible time trip in going from one station
located in the lower universe, say at $l=-l_{1}$, and ending up in the upper
universe station, say at $l=l_{2}$. Following Ref.\cite{MT}, we have located
$l_{1}$ and $l_{2}$ at a value of the radius such that%
\begin{equation}
l_{1}\simeq l_{2}\simeq\frac{10^{4}r_{0}}{\sqrt{1-b^{\prime}\left(
r_{0}\right)  }}%
\end{equation}
that it means $1-b\left(  r\right)  /r\simeq1$ which is in agreement with the
estimates found in Ref.\cite{MT}. The last property we are going to discuss is
the \textquotedblleft total amount\textquotedblright\ of ANEC violating matter
in the spacetime\cite{VKD} which is described by Eq. $\left(  \ref{IV}\right)
$. For the metric $\left(  \ref{Yb(r)0}\right)  $, one obtains%
\begin{align}
I_{V}  &  =\frac{1}{\kappa}\int_{r_{0}}^{r_{0}+\varepsilon}\left(  r-b\left(
r\right)  \right)  \left[  \ln\left(  \frac{e^{2\phi(r)}}{1-\frac{b\left(
r\right)  }{r}}\right)  \right]  ^{\prime}dr=\frac{1}{\kappa}\int_{r_{0}%
}^{r_{0}+\varepsilon}\frac{b\left(  r\right)  }{r}\left(  1+\frac{1}%
{\omega\left(  r\right)  }\right)  dr\label{IV}\\
&  \simeq\frac{1}{\kappa}\left[  \int_{r_{0}}^{r_{0}+\varepsilon}\alpha\mu
r_{0}-\beta\nu r_{0}+1+c\left(  r\right)  \left(  r-r_{0}\right)  \right]  dr,
\end{align}
where we have approximated the expression close to the throat and where we
have defined%
\begin{equation}
c\left(  r\right)  =-\alpha\mu-r_{0}^{-1}+\beta\nu-{\alpha}^{2}{\mu}^{2}%
r_{0}+\alpha\mu r_{0}\beta\,\nu+\alpha{\mu}^{2}r_{0}-\left(  -\alpha\mu
r_{0}+\beta\nu r_{0}\right)  \beta\nu-\beta{\nu}^{2}r_{0}.
\end{equation}
After the integration, we find%
\begin{align}
I_{V}  &  =\frac{1}{\kappa}\left(  \frac{3}{2}{\alpha}r_{0}-\frac{1}{2}%
{\alpha}^{2}r_{0}+{\frac{\nu r_{0}\beta\alpha}{\mu}}-{\frac{\beta\nu r_{0}%
}{\mu}}-{\frac{\alpha}{2\mu}}+\frac{1}{\mu}\right. \nonumber\\
&  \left.  -{\frac{\beta{\nu}^{2}r_{0}}{2{\mu}^{2}}}-{\frac{{\beta}^{2}{\nu
}^{2}r_{0}}{2{\mu}^{2}}}+{\frac{\beta\nu}{2{\mu}^{2}}}-{\frac{1}{2{\mu}%
^{2}r_{0}}}\right)  \underset{\mu r_{0}\gg1}{\simeq}\frac{1}{\kappa}\left(
\frac{3}{2}{\alpha}r_{0}-\frac{1}{2}{\alpha}^{2}r_{0}\right)  ,
\end{align}
and the result is finite. Therefore we can conclude that, in proximity of the
throat the ANEC can be arbitrarily small as it should be.

\end{document}